\documentclass[%
 aip,
 amsmath,amssymb,
 reprint,%
]{revtex4-1}

\usepackage{graphicx}
\usepackage{dcolumn}
\usepackage{bm}

\usepackage[utf8]{inputenc}
\usepackage[T1]{fontenc}
\usepackage{mathptmx}
\usepackage{etoolbox}

\newcommand{\vq}{\mathbf}
\newcommand{\B}{\textrm{B}}
\newcommand{\CE}{\textrm{CE}}

\usepackage{hyperref}

\usepackage{float}
\usepackage{caption}
\usepackage{subcaption}

\usepackage{mathrsfs}
\usepackage{amsfonts}
\usepackage{amsmath}
\usepackage{amsthm}
\usepackage{amssymb}
\usepackage{physics}
\usepackage{dsfont}
\usepackage{esint}

\usepackage[shortlabels]{enumitem}

\usepackage{xcolor}

\makeatletter
\def\@email#1#2{%
 \endgroup
 \patchcmd{\titleblock@produce}
  {\frontmatter@RRAPformat}
  {\frontmatter@RRAPformat{\produce@RRAP{*#1\href{mailto:#2}{#2}}}\frontmatter@RRAPformat}
  {}{}
}%
\makeatother
\begin{document}

\preprint{AIP/123-QED}

\title{Molecular Dynamics Simulation of Hydrodynamic Transport Coefficients in Plasmas}

\author{Briggs Damman}
 \email{bdamman@umich.edu}
\author{Jarett LeVan}
 \email{jarettl@umich.edu}
\author{Scott D. Baalrud}%
 \email{baalrud@umich.edu}
\affiliation{%
    Nuclear Engineering \& Radiological Sciences, University of Michigan, Ann Arbor, Michigan 48109, USA
}%

\date{\today}

\begin{abstract}
Molecular dynamics (MD) simulations are used to calculate transport coefficients in a two-component plasma interacting through a repulsive Coulomb potential. The thermal conductivity, electrical conductivity,  electrothermal coefficient, thermoelectric coefficient, and shear viscosity are computed using the Green-Kubo formalism over a broad range of Coulomb coupling strength, $0.01 \leq \Gamma \leq  140$. Emphasis is placed on testing standard results of the Chapman-Enskog solution in the weakly coupled regime ($\Gamma \ll 1$) using these first-principles simulations. As expected, the results show good agreement for $\Gamma \lesssim 0.1$. However, this agreement is only possible if careful attention is paid to the definitions of linear constitutive relations in each of the theoretical models, a point that is often overlooked. 
For example, the standard Green-Kubo expression for thermal conductivity is a linear combination of thermal conductivity, electrothermal and thermoelectric coefficients computed in the Chapman-Enskog formalism. 
Meaningful results for electrical conductivity are obtained over the full range of coupling strengths explored, but it is shown that potential and virial components of the other transport coefficients diverge in the strongly coupled regime ($\Gamma \gg 1$). In this regime, only the kinetic components of the transport coefficients are meaningful for a classical plasma. 
\end{abstract}

\maketitle


\section{\label{sec:intro} Introduction}
    Accurate models for transport coefficients are necessary for closing  hydrodynamic equations. 
    In plasma physics, this is usually provided by the Chapman-Enskog solution of the Boltzmann equation, which obtains closed-form expressions in terms of plasma conditions.~\cite{Chapman_Cowling,Ferziger_Kaper,Braginskii} 
    A challenge is that these expressions have not been thoroughly validated because making sensitive measurements of transport rates is difficult in plasmas. 
    Although some measurements have been made,~\cite{RynnPF1964,BretzNF1975,WhitePRL1975,TrintchoukPOP2003,HawreliakJPB2004,KuritsynPOP2006,HenchenPOP2019} they tend to have modest precision, are limited to a few coefficients, and scan only a limited range of conditions. 
    Many experimental tests famously do not agree with the predictions of collisional transport theory.~\cite{bohm:1949,HohPOP1960,PowersPF1965,PaulikasPF1962} 
    This is likely because it can be difficult to achieve a truly collisional regime in hot dilute plasmas.
    Here, we apply an alternative method using first-principles molecular dynamics (MD) simulations.\cite{Frenkel02} 
    Because the results provide a formally exact calculation of the hydrodynamic transport coefficients from Newton's equation of motion, they serve to test the assumptions made in plasma kinetic and transport theories. 
    They also provide solutions at conditions of strong coupling, so they can be used to quantify the conditions at which the standard collisional plasma theories are valid, and provide benchmark data to test proposed extensions to strongly coupled conditions.~\cite{LeVan_2025s} 

    
    The simulation results are shown to agree well with the standard plasma theory in the weakly coupled regime $\Gamma_e \lesssim 0.1$. 
    Here, $\Gamma_e$ is the electron Coulomb coupling parameter
    \begin{equation}
    \label{eq:gamma_e}
        \Gamma_e = \frac{e^2}{4\pi\epsilon_{0} a_e k_\B T},
    \end{equation}
     where $e$ is the electric charge, $\epsilon_{0}$ is the permittivity of free space, $a_e = (3/4\pi n_e)^{1/3}$ is the average inter-particle spacing where $n_e$ is the density of ions or electrons (assumed to be equal here), $k_\B$ is Boltzmann's constant, and $T$ is the temperature. 
     The total Coulomb coupling strength has the same definition as Eq.~(\ref{eq:gamma_e}), but where $a$ is based on the total density ($n=n_e+n_i$), so $\Gamma = 2^{1/3}\Gamma_e$. 
     The good agreement between simulations and the standard theory is expected, but it also emphasizes that one must be careful in comparing the results. 
     
    Specifically, in mixtures (electrons and ions) the electrothermal coefficient, thermoelectric coefficient, and thermal conductivity are generally defined differently in kinetic theory and non-equilibrium thermodynamics.\cite{degroot_mazur,LeVan25} 
    The differences come in the way in which one organizes the linear constitutive relations, and whether diffusive contributions are included in the definition of the heat flux.  
    The non-equilibrium thermodynamics definitions are made to preserve Onsager symmetries, whereas the kinetic theory formulations are not.\cite{degroot_mazur,LeVan25} 
    Our MD simulations are based on Green-Kubo relations defined consistently with definitions from the non-equilibrium thermodynamics formulation. 
    A connection between the two definitions is made explicit to facilitate proper comparison. If one is not careful, the standard transport model predictions and MD results, which would otherwise agree, can differ by up to an order of magnitude.
    
    Molecular dynamics simulations have previously been performed extensively for the one-component plasma (OCP) and Yukawa one-component plasma (YOCP) systems.~\cite{Hansen_1975, Bernu_1978, DonkoPRL1998, Donko_2000, SalinPOP2003, Donk_2008, Daligault_2014, Scheiner_2019,OttPRE2015, LeVan_bv} In contrast, two-component ion-electron systems, which are relevant for plasma fluid descriptions, have received comparatively little attention. 
    Because one-component models have no diffusive contribution, they have little relevance to the transport properties of an electron-ion mixture. 
    Specifically, the electrical conductivity, electrothermal coefficient, and thermoelectric coefficients are identically zero in one-component systems. 
    There is a finite thermal conductivity in a one-component system, but this is small compared to the thermal conductivity of the electron-ion system because of the lack of diffusive contributions. 
    The only transport coefficient that is similar in the two systems is the shear viscosity. 
    In fact, it is shown here that the shear viscosity in the two-component system can be accurately obtained from the one-component system.

     Molecular dynamics results are also presented in the strongly coupled regime, extending up to $\Gamma_e = 140$. Although no explicit comparisons are carried out in this work, the data provides a benchmark for classical theories of dense plasmas.\cite{StanekPOP2024,LeVan_2025s,LeePF1984} To avoid the unphysical formation of bound states that would arise in a dense classical system with attractive interactions, all interactions are modeled as purely Coulomb repulsive. In the weakly coupled regime, it is well established that  transport coefficients are independent of the sign of the interaction,\cite{Chapman_Cowling,Ferziger_Kaper,Braginskii} so the results in this regime are completely physical. However, at strong coupling this symmetry breaks down and thus the data is not representative of a real ion-electron plasma.~\cite{ShafferPOP2019,KuzminPRL2002} Instead, the results in this regime are valuable as benchmarks for theories of dense plasmas that attempt to capture the effects of strong correlations. 

     Molecular dynamics simulations of physical dense systems must account for quantum mechanical effects, particularly Pauli blocking, that are responsible for the stability of matter.~\cite{Baus_Hansen_1980,LenardJMP1968} 
     The most common technique is density functional theory molecular dynamics (DFT-MD).~\cite{FrenchPRE2022,WhitePRL2020,DesjarlaisPRE2017,WittePOP2018,SjostromPRL2014} 
     However, these calculations are very computationally expensive, especially at high temperatures. 
     They cannot reach the conditions at which classical plasma theory is expected to be valid in a first-principles way, particularly with dynamic electrons.~\cite{StanekPOP2024,FrenchPRE2022} 
     The few instances of simulations that reach this regime make use of further approximations.~\cite{starrett_2015,WhitePRL2020} 
     Thus, the classical MD simulations have an important role to play in testing plasma theories in the weak coupling regime through the transition to the strong coupling regime, even if they are based on a repulsive potential. 

 The outline of the work is as follows. The relations between transport coefficients derived from kinetic theory and MD are discussed in Sec.~\ref{sec:theory}, the MD simulations are discussed in Sec.~\ref{sec:MD}, a comparison with the CE solution is made in Sec.~\ref{sec:MDresults}, and concluding comments are presented in Sec.~\ref{sec:summary}.

\section{\label{sec:theory} Theory background}
    The macroscopic behavior of sufficiently collisional plasmas can be described by single-fluid magnetohydrodynamic (MHD) equations. Assuming non-relativistic speeds and quasi-neutrality, the conservation of mass, momentum, and energy are written\cite{Ferziger_Kaper,degroot_mazur}
    \begin{subequations}
        \begin{align}
            \frac{1}{\rho} \dv{\rho}{t} &= - \nabla \cdot \vb{V} \label{eq:mhd_mass}, \\
            \rho \dv{\vb{V}}{t} &= - \nabla \cdot \vb{P} \label{eq:mhd_momentum} + \vq j \times \vq B, \\ 
            \label{eq:mhd_energy}
            \rho \dv{u}{t} &= - \nabla \cdot \vb{q} - \vb{P} : \nabla \vb{V} + \vb{j} \cdot \vb{E}', 
        \end{align}
    \end{subequations}
    where $\rho$ is the total mass density, $\vb{V}$ is the center of mass velocity, $\vb{P}$ is the plasma pressure tensor, $\vq B$ is the magnetic field, $\vb{E}' = \vq E + \vq V \times \vq B$ is the electric field in the fluid reference frame, $u$ is the specific internal energy, $\vb{q}$ is the heat flux, and $\vb{j}$ is the diffusive current density. 

    The conservation equations are coupled with Maxwell's equations, which under the same assumptions as above are
    \begin{subequations}
        \begin{align}
            \nabla \times \vq E &= - \frac{\partial \vq B}{\partial t} \\
            \nabla \times \vq B &= \mu_{0} \vq j \\
            \nabla \cdot \vq B &= 0,
        \end{align}
    \end{subequations}
    where $\mu_{0}$ is the vacuum permeability. It is generally desirable to express the MHD equations in terms of the variables $\rho, \vq V, T$, and $\vq B$. Hence, to close the equations, one must specify the equations of state $u(\rho, T)$ and $p(\rho, T)$, and provide constitutive relations that express the fluxes $\vq j$, $\vq q$, and viscous pressure tensor $\hat{\vq P} = \vq P - p\vq I$ in terms of the fluid variables $\rho, \vq V, T$, and $\vq B$. Near equilibrium, the fluxes can be approximated as linear functions of spatial gradients in the fluid variables. The proportionality constants appearing in the linear relations are known as transport coefficients. 
    
    This section presents two frameworks for deriving linear constitutive relations and evaluating transport coefficients. The first approach is based on non-equilibrium thermodynamics and the Green-Kubo relations. It offers a high degree of generality but does not provide a simple means of computing the transport coefficients. Here, the evaluation will be provided from the MD simulation data. The second approach is kinetic theory, which yields algebraic expressions for the transport coefficients, but its validity is limited to dilute (weakly coupled) plasmas. A correspondence between the definitions of transport coefficients in these two frameworks is established, allowing for a consistent comparison between kinetic theory and MD.
    
    \subsection{\label{ssec:noneq_sols} General closure}
    
    A recent tutorial~\cite{LeVan25} reviewed how non-equilibrium thermodynamics can be used to construct general linear constitutive relations for plasmas. In the weakly magnetized limit, all transport coefficients may be represented by scalars, leading to the relations 
        \begin{subequations}
        \label{eq:genlin}
            \begin{align}
                \vb{j} &= \sigma \bigg( \mathbf{E}' - \frac{m_e T}{q_e} \nabla \frac{\mu_e}{T} \bigg) + \varphi \nabla T, \label{eq:GK_lin_curr_dens}\\
                \vb{q} &=  - \lambda \nabla T + \phi \bigg( \mathbf{E}' - \frac{m_e T}{q_e} \nabla \frac{\mu_e}{T} \bigg), \label{eq:GK_lin_heat_flux}\\
                \hat{\vb{P}} &= - \eta \bigg[ \nabla \vb{V} + (\nabla \vb{V})^\texttt{T} - \frac{2}{3}(\nabla \cdot \vb{V})\vb{I}\bigg], \label{eq:GK_lin_pressure}
            \end{align}
        \end{subequations}
        where $\mu_e = \mu_e(\rho, T)$ is the specific electron chemical potential, and $\sigma, \varphi, \lambda, \phi$, and $\eta$ are the electrical conductivity, electrothermal coefficient, thermal conductivity, thermoelectric coefficient, and shear viscosity, respectively.
        
        Equation~(\ref{eq:genlin}) lends itself to the following physical interpretations of the transport coefficients.  The electrical conductivity characterizes the response of the current density to an electric field or chemical potential gradient, and the electrothermal coefficient characterizes the response of the current density to a temperature gradient. Similarly, the thermal conductivity characterizes how a heat flux arises in response to a temperature gradient, and the thermoelectric coefficient characterizes how the heat flux arises in response to an electric field or chemical potential gradient. The shear viscosity characterizes viscous stresses arising from velocity shear.

        In order to facilitate a future comparison with kinetic theory, it is useful to take the ideal gas limit of the linear constitutive relations. In the ideal gas limit,\cite{Reif_1965}
        \begin{equation}
            \mu_{e, \text{ideal}} = -\frac{k_\B T}{m_e} \ln \Bigg[\frac{m_e}{\rho_e} \bigg( \frac{2\pi m_e k_\B T}{h^2}\bigg)^{3/2} \Bigg],
        \end{equation}
        and the linear constitutive relations from Eq.~(\ref{eq:GK_lin_curr_dens}) and (\ref{eq:GK_lin_heat_flux}) become
        \begin{subequations}
            \begin{align}
                \vb{j} &= \sigma\left[\vb{E} - \frac{1}{n q_e} \nabla p + \frac{5k_\B}{2q_e} \nabla T \right] + \varphi \nabla T, \label{eq:GK_exp_lin_curr_dens}\\
                \vb{q} &=  - \lambda \nabla T +\phi\left[\vb{E} - \frac{1}{n q_e} \nabla p + \frac{5k_\B}{2q_e} \nabla T \right] \label{eq:GK_exp_lin_heat_flux}.
            \end{align}
        \end{subequations}
        The linear constitutive relation for the viscous pressure tensor from Eq.~(\ref{eq:GK_lin_pressure}) is unaffected. 
        
        Non-equilibrium thermodynamics does not provide a method to evaluate the transport coefficients. However, one can use the formally exact Green-Kubo relations, which relate transport coefficients to equilibrium fluctuations of the corresponding fluxes.\cite{Hansen_Mcdonald} For a weakly magnetized plasma, these relations take the form\cite{LeVan25}
        \begin{subequations}
        \label{eq:GK}
            \begin{align}
                \sigma &= \frac{V}{3 k_\B T} \int\limits_{0}^{\infty} dt\; \langle \vb{j}(t) \cdot \mathbf{j}(0) \rangle, \label{eq:GK_ec}\\
                \varphi &= - \frac{V}{3 k_\B T^2} \int\limits_{0}^{\infty} dt\; \langle \vb{j}(t) \cdot \mathbf{q}(0) \rangle, \label{eq:GK_et} \\
                \lambda &= \frac{V}{3 k_\B T^2} \int\limits_{0}^{\infty} dt\; \langle \vb{q}(t) \cdot \mathbf{q}(0) \rangle, \label{eq:GK_tc} \\
                \phi &= \frac{V}{3 k_\B T} \int\limits_{0}^{\infty} dt\; \langle \vb{q}(t) \cdot \mathbf{j}(0) \rangle, \label{eq:GK_te} \\
                \eta &= \frac{V}{6 k_\B T} \sum_{i=1}^3 \sum_{j \neq i}^3\int\limits_{0}^\infty dt\; \langle \hat{{P}}_{ij}(t)  \hat{{P}}_{ij}(0) \rangle
            \end{align}
        \end{subequations}
        where $V$ is the volume and $\langle \ldots \rangle$ denotes an ensemble average. To evaluate, it is necessary both to express the fluxes in terms of particle trajectories and to model the microscopic dynamics at equilibrium. The first requirement can be fulfilled using the Irving-Kirkwood procedure.~\cite{Irving_1950,Bearman_1958} One obtains 
        \begin{subequations}
            \label{eq:GK_vectors}
            \begin{align}
                \vb{j} &= \frac{1}{V} \sum\limits_{i=1}^{N} q_i \vb{v}_i, \label{eq:micro_curr_dens} \\
                \vb{P} &= \frac{1}{V} \sum\limits_{i=1}^N \bigg( m_i\vb{v}_i\vb{v}_i + \frac{1}{2}\sum\limits_{j\neq i}^N \vb{r}_{ij} \pdv{\phi_{ij}}{\vb{r}_i}\bigg) \label{eq:micro_pressure} \\
                \vb{q} &= \frac{1}{V} \sum\limits_{i=1}^{N} \bigg[ \underbrace{\vb{v}_{i}\frac{1}{2} m_i \abs{\vb{v}_{i}}^{2}}_{\text{kinetic}} 
                + \underbrace{\frac{1}{2}\vb{v}_i \sum\limits_{j\neq i}^{N} \phi_{ij}}_{\text{potential}} 
                + \underbrace{\frac{1}{2}\sum_{j\neq i}^{N}(\vb{r}_i\cdot \vb{v}_i) \pdv{\phi_{ij}}{\vb{r}_i}}_{\text{virial}} \bigg], \label{eq:micro_heat_flux_comps}
            \end{align}
        \end{subequations}

        where $N$ is the total number of particles; $q_i$, $m_i$, $\vb{r}_i$, and $\mathbf{v}_i$ are the charge, mass, position, and velocity of particle $i$ respectively, and $\phi_{ij}$ and $\vb{r}_{ij}$ are the Coulomb potential and displacement vector between particle $i$ and $j$, respectively.  The viscous pressure tensor is obtained from the total pressure tensor by subtracting the time-averaged value of the latter, i.e.,
        \begin{equation}
            \hat{\vb{P}} = \vb{P} - \langle \vb{P} \rangle_t
        \end{equation}
        In this work, the Green-Kubo relations are evaluated using classical MD simulations. It is important to note that when the fluxes are defined as in Eq.~(\ref{eq:GK_vectors}), the resulting transport coefficients must be plugged into linear constitutive relations of the form Eq.~(\ref{eq:genlin}) to ensure consistent definitions.
        
        For reasons discussed in Sec. \ref{sec:MD}, only the kinetic component of the heat flux is physical for a repulsive Coulomb mixture. Hereafter, when discussing the transport coefficients, only the kinetic heat flux component is considered
        \begin{equation}
            \vb{q}_\text{k} = \frac{1}{V} \sum\limits_{i=1}^{N} \mathbf{v}_{i}\frac{1}{2} m_i \abs{\mathbf{v}_{i}}^{2} \label{eq:GK_kin_comp_heat_flux}.
        \end{equation}
        This will influence the thermal conductivity, thermoelectric coefficient, and electrothermal coefficient at strong coupling. 
        Since the electrical conductivity has no dependence on heat flux, the computed value is the total at any coupling strength. 

    \subsection{\label{ssec:ce_solutions} Kinetic theory closure}
    
        Kinetic theory offers an alternative framework for closing the MHD equations. In particular, the Chapman-Enskog solution of the Boltzmann equation leads to MHD equations with explicit expressions for the transport coefficients. The resulting linear constitutive relations are typically written~\cite{Ferziger_Kaper}
        \begin{subequations}
            \begin{align}
                \vb{j} &= \sigma_{\CE} \bigg[\mathbf{E} - \frac{1}{n q_e} \nabla p \bigg] - \varphi_{\CE} \nabla T  \label{eq:CE_lin_curr_dens}\\
                \vb{q} &= -\lambda_{\CE} \nabla T + \frac{p}{n} \bigg[ \frac{\varphi_{\CE}}{k_\B \sigma_{CE}} + \frac{5}{2q_e}\bigg] \mathbf{j} \label{eq:CE_q} \\
                \hat{\vb{P}} &= - \eta_{\CE} \bigg[ \nabla \vb{V} + (\nabla \vb{V})^\texttt{T} - \frac{2}{3}(\nabla \cdot \vb{V})\vb{I}\bigg] \label{eq:CE_lin_pressure},
            \end{align}
        \end{subequations}
        where the transport coefficients $\sigma_{\CE}$, $\varphi_{\CE}$, $\lambda_{\CE}$, and $\eta_{\CE}$ are not, in general, equivalent to those defined in the previous subsection. It is important to note, however, that the traditional kinetic theory defines microscopic fluxes consistent with those in Eq.~(\ref{eq:GK_vectors}). 
        
        The transport coefficients are written conveniently in terms of the electron Coulomb collision time $\tau_e$, given by
        \begin{equation}
            \tau_e = \frac{3}{2\sqrt{2\pi}} \frac{(4 \pi \epsilon_{0})^2 \sqrt{m_e} (k_\B T)^{3/2}}{n_e q_e^4 \ln{\Lambda}}
        \end{equation}
        where $\ln{\Lambda}$ is the Coulomb logarithm
        \begin{equation}
            \ln{\Lambda} = \ln{\big(\lambda_{\textrm{D}} / r_\textrm{L}\big)}.
        \end{equation}
        Here, $r_\textrm{L} = e^2 / (4 \pi \epsilon_{0} k_\B T)$ is the Landau length and $\lambda_{\textrm{D}} = \sqrt{\epsilon_{0} k_\B T / (e^2 n)}$ is the Debye length. The transport coefficients are then given by~\cite{Ferziger_Kaper, Braginskii}
        \begin{subequations}
            \label{eq:CE_whole}
            \begin{align}
                \sigma_{\CE} &= 1.93 \frac{n_e q_e^2 \tau_e}{2 m_e}, \label{eq:CE_ec} \\
                \varphi_{\CE}& = 0.78 \frac{k_\B n_e q_e \tau_e}{m_e}, \label{eq:CE_et} \\
                \lambda_{\CE} &= 1.02 \frac{n_e k_\B^2 T \tau_e}{m_e}, \label{eq:CE_tc} \\
                \eta_{\CE} &= 0.96 n_i k_\B T \sqrt{\frac{m_i}{2m_e}}\tau_e.
            \end{align}
        \end{subequations}

        To compare with the non-equilibrium thermodynamics form from Eq.~(\ref{eq:GK_exp_lin_heat_flux}), Eq.~(\ref{eq:CE_lin_curr_dens}) and (\ref{eq:CE_q}) are combined to obtain
        \begin{align}
        \label{eq:CE_lin_heat_flux}
                \vb{q} = &- \left( \lambda_{\CE} + T \frac{\varphi^2_{\CE}}{\sigma_{\CE}} + \frac{5k_B T}{2 q_e} \varphi_{\CE}\right) \nabla T \\ \nonumber&+ \left(T \varphi_{\CE} + \frac{5k_\B T}{2 q_e} \sigma_{\CE} \right) \left(\vb E - \frac{1}{n q_e} \nabla p \right). 
       \end{align}
        Then, on rearranging Eq.~(\ref{eq:CE_lin_heat_flux}) to resemble Eq.~(\ref{eq:GK_exp_lin_heat_flux}), and carrying out a similar procedure for the current density, the following relations are identified:
        \begin{subequations}
            \label{eq:GK_CE_whole}
            \begin{align}
                \sigma &= \sigma_{\CE}  = 1.93 \frac{n_e q_e^2 \tau_e}{2 m_e}\label{eq:GK_CE_ec} \\
                \varphi &= -\varphi_{\CE} - \frac{5k_\B}{2q_e}\sigma_{\CE} = -3.19 \frac{k_\B n_e q_e \tau_e}{m_e}\label{eq:GK_CE_et} \\
                 \phi &= T \varphi_{\CE} + \frac{5k_\B T}{2q_e} \sigma_{\CE} = 3.19 \frac{k_\B T n_e q_e \tau_e}{m_e}\label{eq:GK_CE_te} \\
                \lambda &= \lambda_{\CE} + T \frac{\varphi^2_{\CE}}{\sigma_{\CE}} + \frac{5k_\B T}{q_e} \varphi_{\CE} + \frac{25k_\B^2T}{4q_e^2} \sigma_{\CE} \label{eq:GK_CE_tc}\\
                &= 11.58 \frac{n_e k_\B^2 T \tau_e}{m_e} \notag  \\
                \eta &= \eta_{CE} = 0.96 n_i k_\B T \sqrt{\frac{m_i}{2m_e}}\tau_e. \label{eq:GK_CE_shear}
            \end{align}
        \end{subequations}
The relation $\phi = - \varphi T$ is due to Onsager symmetry.\cite{degroot_mazur} 
The coefficients defined in Eq.~(\ref{eq:GK_CE_whole}) are consistent with those defined by the Green-Kubo relations from Eq.~(\ref{eq:GK}). 
These will serve as a basis for comparison with MD simulations, which evaluate the Green-Kubo relations.

\section{\label{sec:MD} MD Simulation Setup}
    
    \begin{figure}
        \centering
        \includegraphics[width=0.9\linewidth]{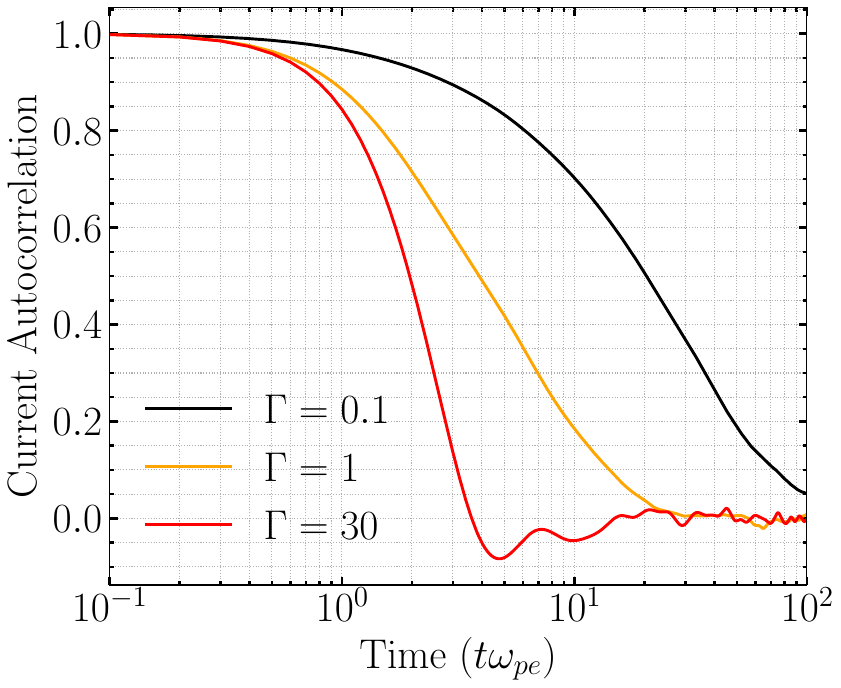}
        \caption{Current autocorrelation functions from simulations at $\Gamma_e = 0.1$ (black), $\Gamma_e = 1$ (yellow), and $\Gamma_e = 30$ (red) that are used to compute the electrical conductivity. The correlation functions are normalized to an initial value of 1.}
        \label{fig:compare_cfs}
    \end{figure}
    
    \begin{figure*}[!ht]
        \centering
        \includegraphics[width=\linewidth]{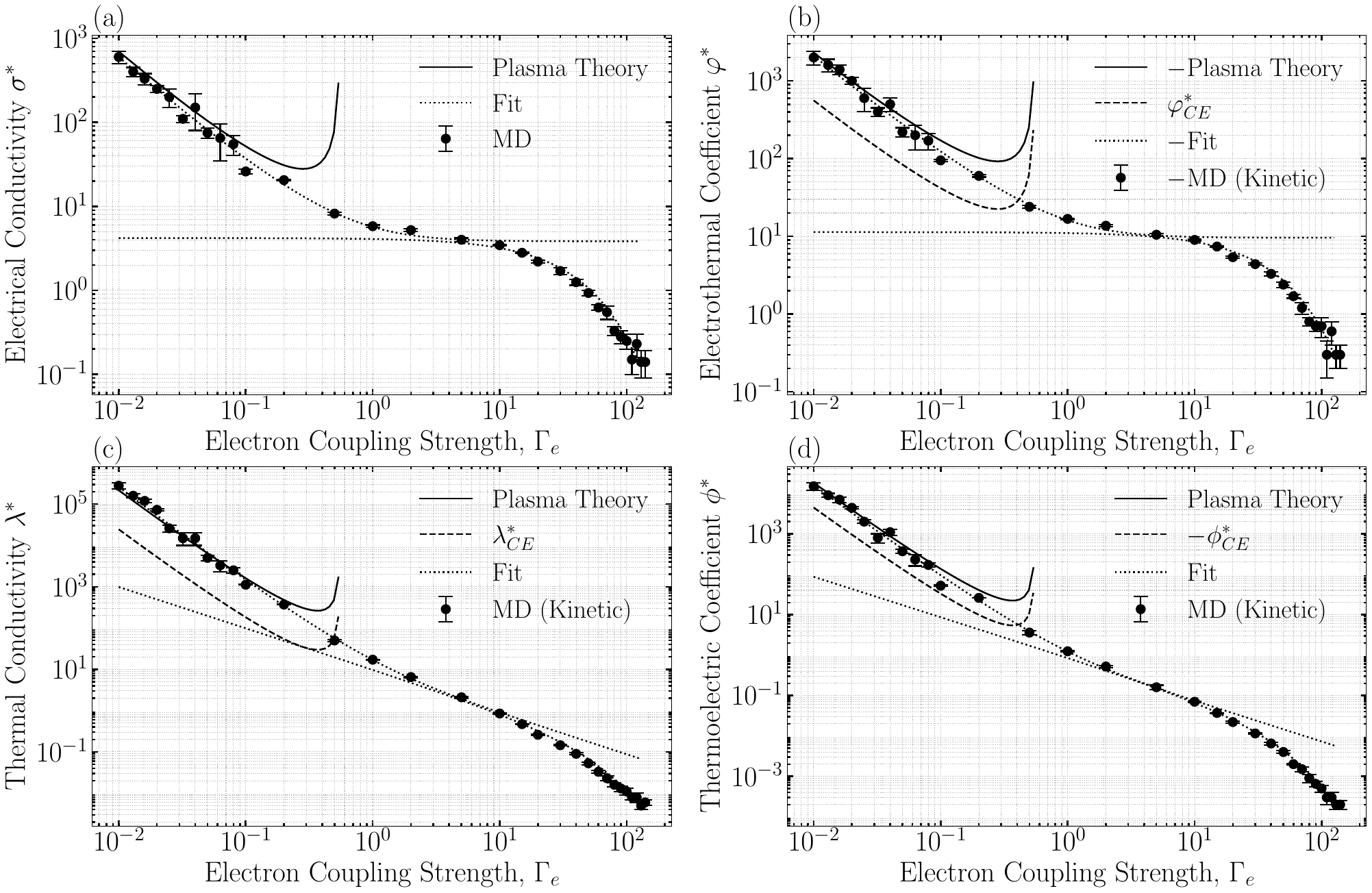}
        \caption{(a) Total electrical conductivity, (b) kinetic electrothermal coefficient, (c) thermal conductivity, and (d) thermoelectric coefficient as a function of $\Gamma_e$. The modified kinetic theory relations from Eq.~(\ref{eq:GK_CE_whole}) (solid lines), and with the definitions from Eq.~(\ref{eq:CE_whole}) (dashed lines) are also plotted for comparison. Dotted lines show the fit formulas from Eqs.~(\ref{eq:xi_wc}) and (\ref{eq:xi_sc}).}
        \label{fig:gvk}
    \end{figure*}
 
    Equilibrium MD simulations were carried out using the LAMMPS simulation platform.~\cite{Plimpton95}  Simulations were run for select values of $\Gamma_e$ in the range $0.01 \leq \Gamma_e \leq 140$.  Initialization at a chosen value of $\Gamma_e$ involved fixing the number of particles and number density, which scales the size of the periodic cubic domain, followed by a 1000 $\omega_{pe}^{-1}$ equilibration phase via a Nos\'{e}-Hoover thermostat to achieve the desired temperature.\cite{Frenkel02}  
    After equilibration, the simulation was run in a microcanonical (energy conserving) ensemble during the data collection phase. 
    Here, $\omega_{pe}^{-1} = (e^2 n_e / (\epsilon_{0} m_e))^{-1/2}$ is the angular electron plasma period, which sets the dimensionless timescale for the simulations. The particles interacted via the Coulomb potential, and the equation of motion was solved using the particle-particle-particle-mesh (P$^3$M) method.\cite{Frenkel02} Table \ref{tab:MD_sim_params} gives the values of the number of particles per species, timestep, and run time for the microcanonical simulation stage for different $\Gamma_e$.  For low $\Gamma_e$, a smaller timestep was used to ensure energy conservation.  However, due to the greater computational difficulty and less of a need to capture longtime oscillations of the CFs for $t \gtrsim 30\omega_{pe}^{-1}$, the total number of plasma periods simulated was less at low $\Gamma_e$.

    \begin{table}[]
        \centering
        \caption{MD Simulation Parameters}
        \begin{tabular}{cccc}
        \hline
            &  $(\Gamma_e \leq .05)$ & $(.05 < \Gamma_e \leq .5)$ & $(\Gamma_e > .5)$\\
        \hline \hline
            Number of particles & 1000 & 5000 & 5000 \\
            Timestep & $0.0001 \omega_{pe}^{-1}$ & $0.001 \omega_{pe}^{-1}$ & $0.01 \omega_{pe}^{-1}$ \\
            Simulation Length & $1\cdot 10^3 \omega_{pe}^{-1}$ & $1\cdot 10^4 \omega_{pe}^{-1}$ & $2\cdot 10^4 \omega_{pe}^{-1}$\\
        \hline
        \end{tabular}
        \label{tab:MD_sim_params}
    \end{table}
    Fluxes were calculated from particle trajectories using the definitions in Eq.~(\ref{eq:GK_vectors}) and used to evaluate the Green-Kubo relations in Eq.~(\ref{eq:GK}). Only the kinetic part of the heat flux was calculated because repulsive Coulomb systems have divergent potential energies. In one-component systems, this fact does not impact the evaluation of transport coefficients because changes in potential energy are well-defined and it is fluctuations which lead to the transport quantities.\cite{Donko_2000,DonkoPRE2004,Scheiner_2019} However, in mixtures, there is a contribution to the heat flux from a diffusion of potential energy, as evidenced by Eq.~(\ref{eq:micro_heat_flux_comps}). This diverges if the potential energy is infinite, making the kinetic component the only physically meaningful contribution. It should be noted that one can obtain finite values for the potential and virial terms from the MD simulation because P$^3$M subtracts a neutralizing background charge distribution. However, the values do not make physical sense and vary depending on the specific distance at which one decides to separate the short-range direct force calculations from the long-range mesh-based force calculations in the P$^3$M algorithm. This is shown in the Appendix. Therefore,  only the kinetic part of the thermal conductivity, electrothermal coefficient, and thermoelectric coefficient will be shown. The electrical conductivity is unaffected, as it is purely kinetic, and the shear viscosity is also unaffected, since it does not include a diffusive contribution. 

    \begin{table}[]
        \begin{center}
\addtolength{\tabcolsep}{3pt}
\begin{tabular}{ccccc}
\hline \hline
$\Gamma$  &  $\sigma^*$  &  $\varphi^*$  &  $\lambda^*$  &  $\phi^*$  \\
\hline
0.01 & $6.0\cross 10^{2}$ & $-2.0\cross 10^{3}$ & $2.8\cross 10^{5}$ & $1.5\cross 10^{4}$ \\
0.013 & $4.0\cross 10^{2}$ & $-1.6\cross 10^{3}$ & $1.6\cross 10^{5}$ & $9.0\cross 10^{3}$ \\
0.016 & $3.3\cross 10^{2}$ & $-1.4\cross 10^{3}$ & $1.2\cross 10^{5}$ & $7.0\cross 10^{3}$ \\
0.02 & $2.5\cross 10^{2}$ & $-1.0\cross 10^{3}$ & $7.3\cross 10^{4}$ & $4.4\cross 10^{3}$ \\
0.025 & $2.0\cross 10^{2}$ & $-6.0\cross 10^{2}$ & $2.6\cross 10^{4}$ & $2.0\cross 10^{3}$ \\
0.032 & $1.1\cross 10^{2}$ & $-4.0\cross 10^{2}$ & $1.5\cross 10^{4}$ & $8.0\cross 10^{2}$ \\
0.04 & $1.5\cross 10^{2}$ & $-5.0\cross 10^{2}$ & $1.5\cross 10^{4}$ & $1.1\cross 10^{3}$ \\
0.05 & $7.5\cross 10^{1}$ & $-2.2\cross 10^{2}$ & $5.0\cross 10^{3}$ & $3.7\cross 10^{2}$ \\
0.063 & $6.5\cross 10^{1}$ & $-2.0\cross 10^{2}$ & $3.3\cross 10^{3}$ & $2.3\cross 10^{2}$ \\
0.08 & $5.5\cross 10^{1}$ & $-1.7\cross 10^{2}$ & $2.5\cross 10^{3}$ & $1.7\cross 10^{2}$ \\
0.1 & $2.6\cross 10^{1}$ & $-9.5\cross 10^{1}$ & $1.1\cross 10^{3}$ & $5.2\cross 10^{1}$ \\
0.2 & $2.0\cross 10^{1}$ & $-6.0\cross 10^{1}$ & $3.7\cross 10^{2}$ & $2.6\cross 10^{1}$ \\
0.5 & $8.2\cross 10^{0}$ & $-2.4\cross 10^{1}$ & $5.0\cross 10^{1}$ & $3.6\cross 10^{0}$ \\
1.0 & $5.8\cross 10^{0}$ & $-1.7\cross 10^{1}$ & $1.7\cross 10^{1}$ & $1.2\cross 10^{0}$ \\
2.0 & $5.2\cross 10^{0}$ & $-1.4\cross 10^{1}$ & $6.5\cross 10^{0}$ & $5.2\cross 10^{-1}$ \\
5.0 & $4.0\cross 10^{0}$ & $-1.0\cross 10^{1}$ & $2.1\cross 10^{0}$ & $1.6\cross 10^{-1}$ \\
10.0 & $3.5\cross 10^{0}$ & $-9.0\cross 10^{0}$ & $8.5\cross 10^{-1}$ & $7.0\cross 10^{-2}$ \\
15.0 & $2.8\cross 10^{0}$ & $-7.4\cross 10^{0}$ & $4.7\cross 10^{-1}$ & $3.7\cross 10^{-2}$ \\
20.0 & $2.2\cross 10^{0}$ & $-5.4\cross 10^{0}$ & $2.6\cross 10^{-1}$ & $2.2\cross 10^{-2}$ \\
30.0 & $1.7\cross 10^{0}$ & $-4.4\cross 10^{0}$ & $1.4\cross 10^{-1}$ & $1.1\cross 10^{-2}$ \\
40.0 & $1.2\cross 10^{0}$ & $-3.3\cross 10^{0}$ & $9.0\cross 10^{-2}$ & $6.5\cross 10^{-3}$ \\
50.0 & $9.3\cross 10^{-1}$ & $-2.4\cross 10^{0}$ & $5.3\cross 10^{-2}$ & $4.0\cross 10^{-3}$ \\
60.0 & $6.2\cross 10^{-1}$ & $-1.7\cross 10^{0}$ & $3.3\cross 10^{-2}$ & $2.0\cross 10^{-3}$ \\
70.0 & $5.5\cross 10^{-1}$ & $-1.2\cross 10^{0}$ & $2.3\cross 10^{-2}$ & $1.5\cross 10^{-3}$ \\
80.0 & $3.3\cross 10^{-1}$ & $-8.0\cross 10^{-1}$ & $1.6\cross 10^{-2}$ & $9.0\cross 10^{-4}$ \\
90.0 & $2.8\cross 10^{-1}$ & $-7.0\cross 10^{-1}$ & $1.3\cross 10^{-2}$ & $6.5\cross 10^{-4}$ \\
100.0 & $2.5\cross 10^{-1}$ & $-7.0\cross 10^{-1}$ & $1.1\cross 10^{-2}$ & $5.0\cross 10^{-4}$ \\
110.0 & $1.5\cross 10^{-1}$ & $-3.0\cross 10^{-1}$ & $8.0\cross 10^{-3}$ & $3.0\cross 10^{-4}$ \\
120.0 & $2.3\cross 10^{-1}$ & $-6.0\cross 10^{-1}$ & $8.0\cross 10^{-3}$ & $3.0\cross 10^{-4}$ \\
130.0 & $1.4\cross 10^{-1}$ & $-3.0\cross 10^{-1}$ & $5.0\cross 10^{-3}$ & $2.0\cross 10^{-4}$ \\
140.0 & $1.4\cross 10^{-1}$ & $-3.0\cross 10^{-1}$ & $6.0\cross 10^{-3}$ & $2.0\cross 10^{-4}$ \\
\hline \hline
\end{tabular}
\addtolength{\tabcolsep}{-3pt}
\end{center}

        \caption{Dimensionless kinetic components of the transport coefficients, calculated using the Green-Kubo formalism.}
        \label{tab:MD_data}
    \end{table}

    To evaluate the Green-Kubo relations from MD simulations, time must be discretized and cutoff after a large number of timesteps $N_L$, by when the correlation function has decayed to zero. The ensemble average is replaced by a time average, so one obtains a time-series with $N_T \gg N_L$ timesteps that is used to average $N_T - N_L$ correlation functions. Hence, for a generalized coefficient $L_{ik}$ (related to the previously defined transport coefficients by a sign and temperature factor), the Green-Kubo relations are written
    \begin{equation}
        L_{ik} = \frac{V}{k_\B} \frac{\Delta t}{N_T - N_L + 1} \sum\limits_{\tau_L = 0}^{N_L}\sum\limits_{\tau=0}^{N_T - N_L} \mathcal{J}_{i}(\tau + \tau_L) \mathcal{J}_k(\tau) \label{eq:gen_GK}
    \end{equation}
    where $\Delta t$ is the length of a timestep and $t_N$ is the total number of timesteps. $L_{ik}$ is the generalized coefficient resulting from the fluxes $\mathcal{J}_i$ and $\mathcal{J}_k$. The MD data calculated from this method can be seen in Table~(\ref{tab:MD_data}).
    
    As an example of what the correlation functions look like, the normalized correlation functions for the electrical conductivity at $\Gamma_e = 0.1, 1,$ and $30$ is presented in Figure~\ref{fig:compare_cfs}. As $\Gamma_e$ increases, the correlation functions decay faster and begin to exhibit oscillations. This trend is typical of correlation functions in the transition from weak to strong coupling and is also seen in the other coefficients.

\section{MD Simulation Results\label{sec:MDresults}}

\subsection{\label{sec:elec_cond} Electrical Conductivity}
    
    Molecular dynamics results for the electrical conductivity and the Champan-Enskog result from Eq.~(\ref{eq:CE_ec}) are shown in Fig.~\ref{fig:gvk}a. Data is presented in terms of the dimensionless conductivity $\sigma^*$, defined by
    \begin{equation}
        \sigma^* = \frac{\sigma}{\epsilon_{0} \omega_{pe}}.
    \end{equation}
    Recall that there is no potential contribution to the conductivity, so results represent the total electrical conductivity for all values of $\Gamma_e$.

    For $\Gamma_e < 0.1$, the electrical conductivity is shown to decrease with increasing $\Gamma_e$. The traditional plasma theory result from Eq.~(\ref{eq:GK_CE_ec}), which predicts
    \begin{equation}
        \sigma \propto T^{3/2}/ \ln \Lambda,
    \end{equation}
    shows excellent agreement with MD in this regime. 
    Although electrical conductivity is an important quantity in plasma physics, it is difficult to measure in an experiment, especially for a broad range of plasma conditions. 
    The comparison with first-principles MD simulations is valuable because it confirms that the many approximations that go into the Boltzmann kinetic equation, binary collision description, and Chapman-Enskog solution, are accurate in the regime in which they are expected to be $\Gamma_e \lesssim 0.1$.

    For $0.1 < \Gamma_e < 20$, the MD data transitions toward a plateau at $\sigma^* \approx 5$, before again scaling steeply with an inverse power of $\Gamma_e$ when $\Gamma_e \gtrsim 20$. As expected, traditional plasma theory fails to capture this because it is based on assumptions associated with weak correlations, and the the Coulomb logarithm diverges as $\Gamma_e$ approaches 1. 
    The qualitative change in scaling when $\Gamma_e > 20$, indicates a transition to a liquid-like state, as has been observed in other transport coefficients computed from MD for the one-component plasma.~\cite{DaligaultPRL2006,DonkoPRL2002} 
    Specifically, the trend in the electrical conductivity with increasing $\Gamma_e$ tracks similarly to what was observed for the self-diffusion coefficient of the one-component plasma,~\cite{DaligaultPRL2006,Hansen_1975,OhtaPOP2000} or the interdiffusion in a mixture.~\cite{DaligaultPRL2012} 
    This is expected since electrical conductivity is essentially a process of the interdiffusion of electrons and ions.

%

\subsection{\label{sec:et_te_coeff} Electrothermal and Thermoelectric Coefficients}
    
    Molecular dynamics results for the kinetic part of the electrothermal coefficient are shown in Fig.~\ref{fig:gvk}b, in terms of the dimensionless coefficient 
    \begin{equation}
        \varphi^* = \frac{\varphi}{k_\B n_e e / m_e \omega_{pe}} .
    \end{equation}
    The results show a similar trend to that of the electrical conductivity. Specifically, the coefficient decreases with increasing $\Gamma_e$ for $\Gamma_e < 0.1$, transitions to a plateau for $0.1 < \Gamma_e < 20$, and decreases for $\Gamma_e > 20$. 

    To compare with the traditional plasma theory, Eq.~(\ref{eq:GK_CE_et}) is used to account for the differing definitions between this and the method from non-equilibrium thermodynamics. The coefficient associated with consistent definitions is shown by the solid line in Fig.~\ref{fig:gvk}, and again excellent agreement with the MD is observed at weak coupling $\Gamma_e < 0.1$. For comparison, the coefficient $\varphi_{\CE}^*$, from Eq.~(\ref{eq:CE_et}) is also shown (dashed line). 
    Although this is not expected to compare well with the MD data, because of the inconsistent definitions of the linear constitutive relation in each formulation, it is shown to emphasize the importance of being careful on how the linear constitutive relations are defined to accurately model transport properties. 
    In particular, this coefficient differs by a sign and roughly an order of magnitude from the MD data. 
    This highlights that if one were to take the electrothermal coefficient from one formulation of the linear constitutive relations, but use it in another, the results can be widely inaccurate and unphysical. 
    This is especially emphasized because the electrothermal effects are often ignored in plasmas, though they are intimately related to the thermal conductivity and the definition of Fourier's heat law; as well as in Ohm's law. 

    Results for the thermoelectric coefficient ($\phi^* = \phi / (e \omega_{pe} / a_e)$) are shown in Fig.~\ref{fig:gvk}d. Since it is related to the electrothermal coefficient by the Onsager relation $\phi = -\varphi T$, similar agreement between the MD and CE results are observed.     

\subsection{\label{sec:ther_cond} Thermal Conductivity}
    \begin{figure}
        \centering
        \includegraphics[width=\linewidth]{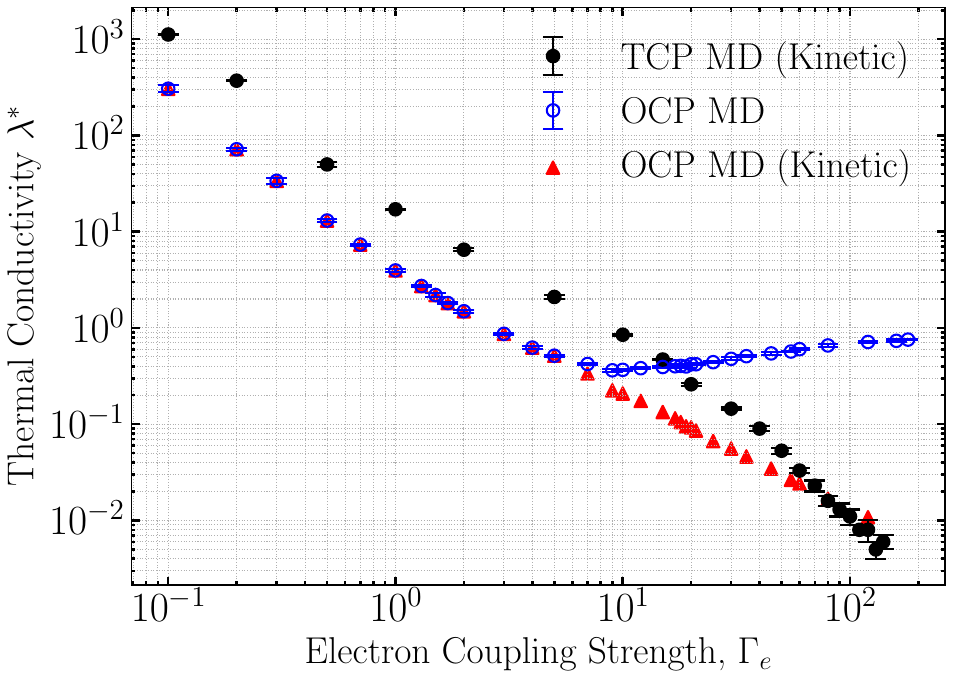}
        \caption{Thermal conductivity computed from MD as a function of $\Gamma_e$ for: the kinetic component in a  two-component plasma (black circles), total in a one-component plasma (blue circles) and kinetic component in a one-component plasma (red triangles). Here the OCP data is from Ref.~\onlinecite{Scheiner_2019}.} 
        \label{fig:tc_ocp_vs_tcp}
    \end{figure}
    Molecular dynamics results for the kinetic part of the thermal conductivity are shown in Fig.~\ref{fig:gvk}c in terms of the dimensionless parameter 
    \begin{equation}
        \lambda^* = \frac{\lambda} {n_e \omega_{pe} k_\B a_e^2}.
    \end{equation}
    The MD data shows that $\lambda^*$ decreases for increasing $\Gamma_e$ across the entire studied range. Consistent with the previous sections, the CE result agrees very well with MD for $\Gamma_e < 0.1$ when using the appropriate relation from Eq.~(\ref{eq:GK_CE_tc}). Also plotted is the coefficient $\lambda_{\CE}^*$, which is an order of magnitude smaller than $\lambda^*$, demonstrating the importance of being consistent in the definitions of transport coefficients and linear constitutive relations.    

    Thermal conductivity has previously been computed from MD for the one-component plasma.~\cite{Bernu_1978,DonkoPRL1998,DonkoPRE2004,Scheiner_2019} 
    However, thermal conduction in a one-component system has a fundamentally different character than in a two-component system, so the results differ dramatically; see Fig.~\ref{fig:tc_ocp_vs_tcp}. 
    Specifically, there is no electrical current in a one-component system ($\vb{j} =0$) as a consequence of momentum conservation. 
    This renders the electrical conductivity, thermoelectric, and electrothermal coefficients to all be zero; see Eq.~(\ref{eq:GK}). 
    As a result, each of these diffusive contributions in Eq.~(\ref{eq:GK_CE_tc}) vanish, along with the thermal diffusion contribution to $\lambda_{\CE}$, resulting in the total thermal conductivity in the one-component system to be equivalent to the reduced thermal conductivity $\lambda_\CE^\prime$, see Ref.~\onlinecite{Ferziger_Kaper}. 
    The results for a one-component system~\cite{Bernu_1978,DonkoPRL1998,DonkoPRE2004,Scheiner_2019} are shown in Fig.~\ref{fig:tc_ocp_vs_tcp} alongside those for the two-component system. The values for the one-component system are nearly an order of magnitude less than in the two-component system at weak coupling. 
    This emphasizes the importance of diffusive processes in the two-component system. 

    It should also be emphasized that MD results for the two-component system contain only the kinetic contribution to the thermal conductivity. 
    As was shown in the one-component plasma,\cite{Bernu_1978,DonkoPRL1998,DonkoPRE2004,Scheiner_2019} it is expected that the virial and potential contributions to the heat flux [see Eq.~(\ref{eq:GK_exp_lin_heat_flux})] overcome the kinetic contribution at strong coupling and change the scaling such that the thermal conductivity increases with $\Gamma_e$; see Fig.~\ref{fig:tc_ocp_vs_tcp}. 
    In the one-component system, this leads to a minimum of the thermal conductivity for $\Gamma_e \approx 17$.~\cite{Scheiner_2019} 
    It should be expected that a similar behavior will be observed for a physical strongly coupled plasma, such as a dense plasma, but this cannot be observed in the repulsive electron-ion system modeled here because the virial and potential terms diverge; as described in the Appendix. 


\subsection{\label{sec:shear_viscosity} Shear Viscosity}
    \begin{figure}
        \centering
        \includegraphics[width=\linewidth]{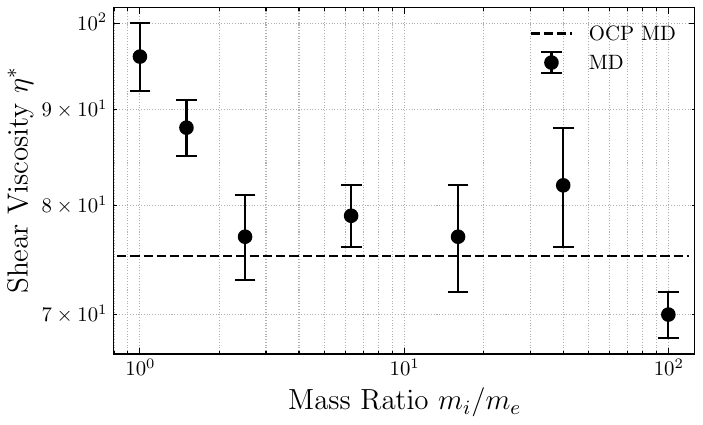}
        \caption{Shear viscosity plotted as a function of mass ratio at~$\Gamma_e = 0.1$. The modified kinetic theory relation from Eq.~(\ref{eq:GK_CE_shear}) (solid line), and OCP MD value (dashed line) are also plotted.}
        \label{fig:shear_vs_mr}
    \end{figure}
    \begin{figure}
        \centering
        \includegraphics[width=\linewidth]{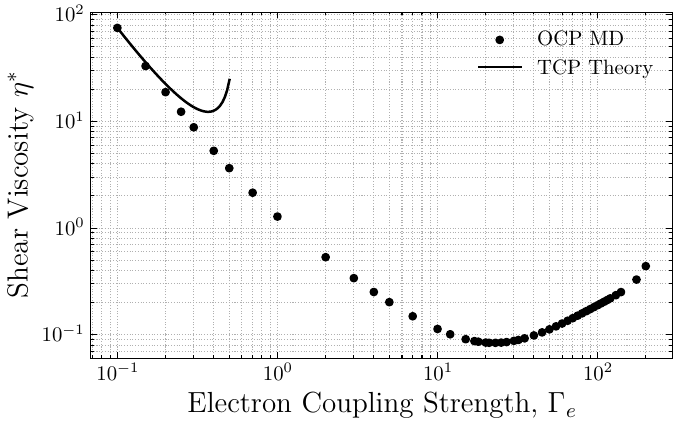}
        \caption{OCP shear viscosity plotted as a function of $\Gamma_e$. The modified kinetic theory relation from Eq.~(\ref{eq:GK_CE_shear}) is also plotted.}
        \label{fig:shear_vs_gamma}
    \end{figure}

    Computing shear viscosity from MD simulations is much more challenging in a two-component system than the conductivity or thermoelectric coefficients. 
    This is because ions carry most of the momentum, so the viscosity is associated with ion motion, which occurs at a much longer timescale than the electron-dominated processes. 
    Resolving both the electron motion with a small timestep, and the long-time decay of the shear stress autocorrelation function associated with ion motion is impractical for a large set of conditions. 
    However, since electrons contribute little to the process, it justifies using a one-component model to simulate shear viscosity.

    Figure~(\ref{fig:shear_vs_mr}) shows MD results from the two-component system at $\Gamma_e = 0.1$ as a function of mass ratio. 
    Here, shear viscosity is presented in terms of the dimensionless parameter
    \begin{equation}
        \eta^* = \frac{\eta}{m_i n_i a_i^2 \omega_{pi}} \label{eq:shear_norm}
    \end{equation}
    where~$\omega_{pi} = (e^2 n_i / (\epsilon_0 m_i))^{1/2}$ is the angular ion plasma frequency.
    This shows that as the mass ratio increases from 1 to approximately 10, the result asymptotes to a slightly smaller ($\sim 15\%$) value. 
    This signifies the fact that both electrons and ions contribute at unity mass ratio, but only ions do at a large mass ratio. 
    It is expected that at a large mass ratio, electrons contribute only by participating in screening ion interactions, which is a second-order effect. 
    This notion is supported by the fact that the MD results asymptote to the value obtained from an OCP simulation at the same density as the ion density in the two-component simulation. 
    The seemingly larger disagreement at the highest mass ratios is likely because these simulations become harder to resolve due to the slower decay of the shear stress autocorrelation function. 

    Using this as a justification, Fig.~\ref{fig:shear_vs_gamma} presents the shear viscosity coefficient from the one-component plasma previously computed in Ref.~\onlinecite{Daligault_2014}. 
    This shows good agreement with the prediction from Eq.~(\ref{eq:GK_CE_shear}) when the density of the one-component simulation is interpreted as the ion density in the two-component system. 
    As has been discussed previously,~\cite{Daligault_2014} shear viscosity has a minimum at $\Gamma_e \approx 17$, corresponding to the conditions that the potential component of the shear stress becomes larger than the kinetic component; see Eq.~(\ref{eq:micro_pressure}). 
    This also signifies a transition to a liquid-like regime in the range $20\lesssim \Gamma_e \lesssim 170$.~\cite{DaligaultPRL2006}



    \subsection{\label{sec:fits}  Fit Formulas}
    
    Formulas fit to the electrical conductivity, electrothermal coefficient, thermal conductivity, and thermoelectric coefficient are provided for ease of using the MD results. 
 At low $\Gamma_e = [\Gamma_{\text{min}}, 10]$, each coefficient is fit to its respective theory equation in Eq.~(\ref{eq:GK_CE_whole}) with an added multiplicative factor $\alpha_\xi$ and the modified Coulomb logarithm 
    \begin{equation}
    \label{eq:gen_ln}
        \ln \Lambda = \ln(1 + C_\xi \frac{\lambda_{D}}{r_L}).
    \end{equation} 
    Explicitly, for the small $\Gamma_e$ values
    \begin{equation}
    \label{eq:xi_wc}
        \xi_{\text{WC,Fit}}^* = \alpha_\xi \xi_{\text{WC}}^* \frac{\ln(\lambda_{D}/r_L)}{\ln(1 + C_\xi \frac{\lambda_{D}}{r_L})}
    \end{equation}
    where $\alpha_\xi$ and $C_\xi$ are the tunable parameters, and $\xi_{\text{WC}}^*$ is the corresponding transport coefficient from Eq.~(\ref{eq:GK_CE_whole}). 
    This essentially just replaces the Coulomb logarithm in the traditional formula with Eq.~(\ref{eq:gen_ln}), and scales the magnitude to the fit the MD data. 
    For large $\Gamma_e$ values $\Gamma_e = [2, \Gamma_{\text{max}}]$, the coefficients are fit to
    \begin{equation}
    \label{eq:xi_sc}
        \xi_{\text{SC,Fit}}^* = A_\xi e^{-B_\xi \Gamma_e} \begin{cases}
            1, &\xi = \sigma, \varphi\\
            \frac{1}{\Gamma_e}, &\xi = \lambda, \phi
        \end{cases} 
    \end{equation}
    where $A_\xi$ and $B_\xi$ are the tuneable parameters. 
    This form is motivated by the Eyring model in liquids.\cite{Hansen_Mcdonald} 
    Both of these fit formulas are based on the application to diffusion in a two-component ion plasma from Ref.~\onlinecite{DaligaultPRL2012}.

        Results for the fits are shown in Figure~(\ref{fig:gvk}) with the fit parameters provided in Table~(\ref{tab:MD_fit_params}). 
    Recall that these represent a fit of the kinetic components of the transport coefficients only.
The fit formulas are able to accurately represent the trends in the data in all cases. 
This emphasizes that each of the electrical transport processes is predominately a diffusive process, as the fit formulas were developed to describe diffusion.~\cite{DaligaultPRL2012}
    Although we do not repeat it here, a fit of the OCP shear viscosity was previously provided in Ref.~\onlinecite{Daligault_2014}.

\section{\label{sec:summary} Summary}
    \begin{table}
        \centering
        \addtolength{\tabcolsep}{5pt}
        \begin{tabular}{c c c c c}
            \hline \hline
             & $\alpha_\xi$ & $C_\xi$ & $A_\xi$ & $B_\xi$ \\
             \hline
             $\sigma^*$ & 1.00 & 2.68 & 4.20 & 0.027\\
             $\varphi^*$ & 1.12 & 3.95 & 11.34 & 0.029 \\
             $\lambda^*$ & 1.55 & 5.65 & 9.87 & 0.022 \\
             $\phi^*$ & 1.08 & 4.01 & 0.87 & 0.028 \\
            \hline \hline
        \end{tabular}
        \addtolength{\tabcolsep}{-5pt}
        \caption{Fitting parameters for the dimensionless data provided in Table~(\ref{tab:MD_data}). The functions used to fit are described in Sec.~(\ref{sec:fits}).}
        \label{tab:MD_fit_params}
    \end{table}
    In this work, MD simulations were used to evaluate transport coefficients in a two-component plasma interacting through a repulsive Coulomb potential. In particular, the thermal conductivity, electrical conductivity, electrothermal coefficient, thermoelectric coefficient, and shear viscosity were computed using the Green-Kubo formalism. Results were used to benchmark the Chapman-Enskog solution of the Boltzmann equation, and good agreement was found up to a Coulomb coupling strength of $\Gamma_e \lesssim 0.1$. It was shown that agreement is only possible if one pays careful attention to the differing definitions of transport coefficients in kinetic theory and MD. Data in the strongly coupled regime, up to $\Gamma_e = 140$, was also provided to serve as a benchmark for classical theories of strongly coupled plasmas. 
    Here, only the kinetic contributions were used to evaluate the heat flux. 

\begin{acknowledgments}
    This work was supported by NSF grant no. PHY-2205506, the NNSA Stewardship Science Academic Programs under DOE Cooperative Agreement DE-NA0004148, and the DOE NNSA Stockpile Stewardship Graduate Fellowship through cooperative agreement DE-NA0004185.
\end{acknowledgments}

\section*{Author declarations}
The authors have no conflicts to disclose.

\section*{Data Availability Statement}
The data that supports the findings of this study are available within the article.

\appendix 

\section{\label{app:diff_in_approx} Divergence of Potential and Virial Components}
    \begin{figure*}
        \includegraphics[width=\linewidth]{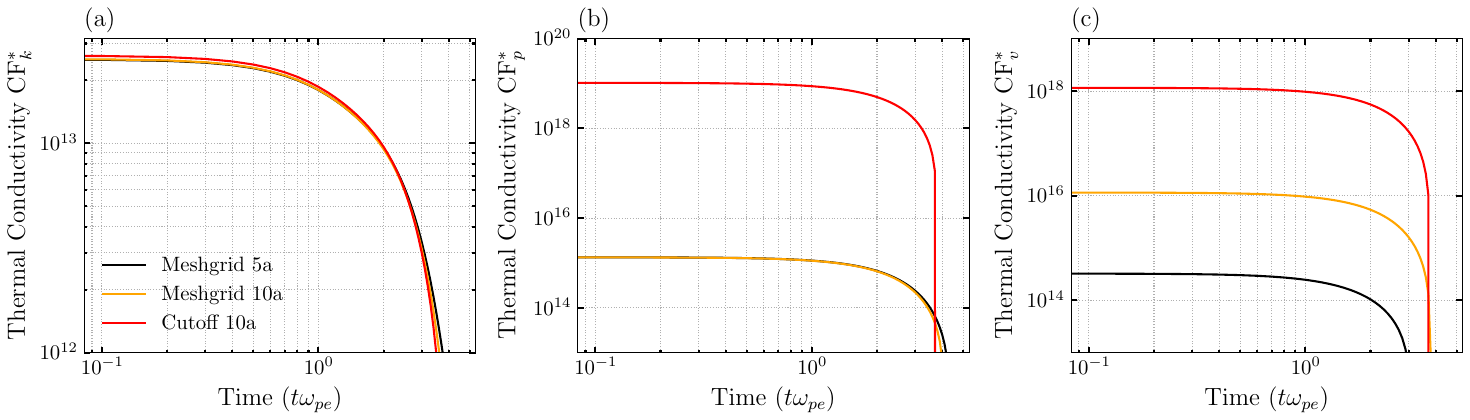}
        \caption{Components of the thermal conductivity's correlation functions plotted at $\Gamma_e=20$ using different methods of potential approximation. The kinetic components are plotted in (a), potential in (b), and virial in (c). The Meshgrid lines correspond to a LAMMPS simulation using P$^3$M at a distance of $5a_e$ (black line) and $10a_e$ (yellow line). The red line corresponds to an MD simulation with a hard cutoff for the potential at $10a_e$.}
        \label{fig:appendix}
    \end{figure*}

    During the process of simulating the two-component plasma, it was noticed that using different cutoff distances or methods of approximating the Coulomb potential resulted in different values in the potential and virial components of the heat flux (Equation \ref{eq:micro_heat_flux_comps}). Components of the transport coefficient's correlation functions at $\Gamma_e=20$ using different methods of potential approximation are plotted in Figure \ref{fig:appendix}. The kinetic components are plotted in (a), potential in (b), and virial in (c). Figure \ref{fig:appendix} (a) shows that the kinetic component is unaffected by the change in potential approximation. However, the potential and virial components, given in Figures \ref{fig:appendix} (b) and \ref{fig:appendix} (c), are shown to differ significantly. Physically, this occurs because in a repulsive Coulomb system, the potential energy of a particle diverges in the thermodynamic limit. Since the simulations use P$^3$M, a finite but meaningless value is given for the potential energy. 
    
    We note that in a one-component Coulomb system, this is not an issue. Consider the potential part of the heat flux 
    \begin{equation}
        \vq q_{\mathrm{pot}} = \frac{1}{V}\sum_{i=1}^N \vq v_i \phi_{i}
    \end{equation}
where $\phi_i = \sum_{j\neq i}^N \phi_{ij} / 2$. In the center-of-mass frame, momentum conservation allows one to write
\begin{equation}
    \vq q_{\mathrm{pot}} = \frac{1}{V} \sum_{i=1}^{N-1} \vq v_i (\phi_{i} - \phi_N),
\end{equation}
so that it is only potential energy differences between two particles which are relevant. Thus, $\vq q_{\mathrm{pot}}$ is not sensitive to differing methods of approximating Coulomb interactions in a one-component system. 

In a two-component system, one does not have the condition $\sum_i \vq v_i = 0$ in the center-of-mass frame, so the magnitude of the potential energy (as opposed to changes in potential energy) is relevant. This is due to the existence of diffusion - heat can transfer through a surface due to a net flux of particles, and since these particles carry divergent potential energy, the potential part of the heat flux is divergent. Meaningful results for the potential and virial parts of the heat flux in a Coulomb mixture can only be obtained if one uses attractive interactions so that the potential energy is finite.  
\newpage
\bibliography{references}

@ARTICLE{Braginskii,
       author = {{Braginskii}, S.~I.},
        title = "{Transport Processes in a Plasma}",
      journal = {Reviews of Plasma Physics},
         year = 1965,
        month = jan,
       volume = {1},
        pages = {205},
  
}

@article{Bearman_1958,
    author = {Bearman, Richard J. and Kirkwood, John G.},
    title = {Statistical Mechanics of Transport Processes. XI. Equations of Transport in Multicomponent Systems},
    journal = {The Journal of Chemical Physics},
    volume = {28},
    number = {1},
    pages = {136-145},
    year = {1958},
    month = {01},
    issn = {0021-9606},
    doi = {10.1063/1.1744056},
}

@article{Irving_1950,
    author = {Irving, J. H. and Kirkwood, John G.},
    title = {The Statistical Mechanical Theory of Transport Processes. IV. The Equations of Hydrodynamics},
    journal = {The Journal of Chemical Physics},
    volume = {18},
    number = {6},
    pages = {817-829},
    year = {1950},
    month = {06},
    issn = {0021-9606},
    doi = {10.1063/1.1747782},
}

@article{LeVan_bv,
  title = {Intrinsic bulk viscosity of the one-component plasma},
  author = {LeVan, Jarett and Baalrud, Scott D.},
  journal = {Phys. Rev. E},
  volume = {111},
  issue = {1},
  pages = {015202},
  numpages = {11},
  year = {2025},
  month = {Jan},
  publisher = {American Physical Society},
  doi = {10.1103/PhysRevE.111.015202},
  url = {https://link.aps.org/doi/10.1103/PhysRevE.111.015202}
}

@article{Daligault_2014,
  title = {Determination of the shear viscosity of the one-component plasma},
  author = {Daligault, J\'er\^ome and Rasmussen, Kim \O{}. and Baalrud, Scott D.},
  journal = {Phys. Rev. E},
  volume = {90},
  issue = {3},
  pages = {033105},
  numpages = {11},
  year = {2014},
  month = {Sep},
  publisher = {American Physical Society},
  doi = {10.1103/PhysRevE.90.033105},
  url = {https://link.aps.org/doi/10.1103/PhysRevE.90.033105}
}

@article{Scheiner_2019,
  title = {Testing thermal conductivity models with equilibrium molecular dynamics simulations of the one-component plasma},
  author = {Scheiner, Brett and Baalrud, Scott D.},
  journal = {Phys. Rev. E},
  volume = {100},
  issue = {4},
  pages = {043206},
  numpages = {10},
  year = {2019},
  month = {Oct},
  publisher = {American Physical Society},
  doi = {10.1103/PhysRevE.100.043206},
  url = {https://link.aps.org/doi/10.1103/PhysRevE.100.043206}
}

@article{Donk_2008,
  title = {Shear viscosity of strongly coupled Yukawa liquids},
  author = {Donk\'o, Z. and Hartmann, P.},
  journal = {Phys. Rev. E},
  volume = {78},
  issue = {2},
  pages = {026408},
  numpages = {8},
  year = {2008},
  month = {Aug},
  publisher = {American Physical Society},
  doi = {10.1103/PhysRevE.78.026408},
}

@article{Donko_2000,
    author = {Donkó, Z. and Nyiri, B.},
    title = {Molecular dynamics calculation of the thermal conductivity and shear viscosity of the classical one-component plasma},
    journal = {Physics of Plasmas},
    volume = {7},
    number = {1},
    pages = {45-50},
    year = {2000},
    month = {01},
    issn = {1070-664X},
    doi = {10.1063/1.873824},
}

@article{Bernu_1978,
  title = {Transport coefficients of the classical one-component plasma},
  author = {Bernu, B. and Vieillefosse, P.},
  journal = {Phys. Rev. A},
  volume = {18},
  issue = {5},
  pages = {2345--2355},
  numpages = {0},
  year = {1978},
  month = {Nov},
  publisher = {American Physical Society},
  doi = {10.1103/PhysRevA.18.2345},
}

@article{Hansen_1975,
  title = {Statistical mechanics of dense ionized matter. III. Dynamical properties of the classical one-component plasma},
  author = {Hansen, J. -P. and McDonald, I. R. and Pollock, E. L.},
  journal = {Phys. Rev. A},
  volume = {11},
  issue = {3},
  pages = {1025--1039},
  numpages = {0},
  year = {1975},
  month = {Mar},
  publisher = {American Physical Society},
  doi = {10.1103/PhysRevA.11.1025},
}

@ARTICLE{Plimpton95,
  title     = {\href{https://doi.org/10.1006/jcph.1995.1039}{Fast Parallel Algorithms for Short-Range Molecular Dynamics}},
  author    = "Plimpton, Steve",
  journal   = "J. Comput. Phys.",
  publisher = "Elsevier BV",
  volume    =  117,
  number    =  1,
  pages     = "1--19",
  year      =  1995,
}

@article{LeVan25,
  title = {Foundations of magnetohydrodynamics},
  volume = {32},
  ISSN = {1089-7674},
  url = {http://dx.doi.org/10.1063/5.0274784},
  DOI = {10.1063/5.0274784},
  number = {7},
  journal = {Physics of Plasmas},
  publisher = {AIP Publishing},
  author = {LeVan,  Jarett and Baalrud,  Scott D.},
  year = {2025},
  month = jul 
}

@ARTICLE{Frenkel02,
    author = {Frenkel, Daan and Smit, Berend},
    year = {2002},
    title = {Understanding molecular simulation: from algorithms to applications. 2nd ed},
    journal = {Academic Press},
    doi = {10.1063/1.881812}
}

@book{Hansen_Mcdonald,
  title={Theory of simple liquids: with applications to soft matter},
  author={Hansen, Jean-Pierre and McDonald, Ian Ranald},
  year={2013},
  publisher={Academic press}
}

@article{DonkoPRL1998,
  title = {Thermal Conductivity of the Classical Electron One-Component Plasma},
  author = {Donk\'o, Z. and Ny\'{\i}ri, B. and Szalai, L. and Holl\'o, S.},
  journal = {Phys. Rev. Lett.},
  volume = {81},
  issue = {8},
  pages = {1622--1625},
  numpages = {0},
  year = {1998},
  month = {Aug},
  publisher = {American Physical Society},
  doi = {10.1103/PhysRevLett.81.1622},
}

@article{ShafferPOP2019,
    author = {Shaffer, Nathaniel R. and Baalrud, Scott D.},
    title = {The Barkas effect in plasma transport},
    journal = {Physics of Plasmas},
    volume = {26},
    number = {3},
    pages = {032110},
    year = {2019},
    month = {03},
    issn = {1070-664X},
    doi = {10.1063/1.5089140},
    url = {https://doi.org/10.1063/1.5089140},
}

@article{DaligaultPRL2006,
  title = {Liquid-State Properties of a One-Component Plasma},
  author = {Daligault, J\'er\^ome},
  journal = {Phys. Rev. Lett.},
  volume = {96},
  issue = {6},
  pages = {065003},
  numpages = {4},
  year = {2006},
  month = {Feb},
  publisher = {American Physical Society},
  doi = {10.1103/PhysRevLett.96.065003},
  url = {https://link.aps.org/doi/10.1103/PhysRevLett.96.065003}
}

@article{DonkoPRL2002,
  title = {Caging of Particles in One-Component Plasmas},
  author = {Donk\'o, Z. and Kalman, G. J. and Golden, K. I.},
  journal = {Phys. Rev. Lett.},
  volume = {88},
  issue = {22},
  pages = {225001},
  numpages = {4},
  year = {2002},
  month = {May},
  publisher = {American Physical Society},
  doi = {10.1103/PhysRevLett.88.225001},
  url = {https://link.aps.org/doi/10.1103/PhysRevLett.88.225001}
}

@article{DaligaultPRL2012,
  title = {Diffusion in Ionic Mixtures across Coupling Regimes},
  author = {Daligault, J\'er\^ome},
  journal = {Phys. Rev. Lett.},
  volume = {108},
  issue = {22},
  pages = {225004},
  numpages = {5},
  year = {2012},
  month = {May},
  publisher = {American Physical Society},
  doi = {10.1103/PhysRevLett.108.225004},
  url = {https://link.aps.org/doi/10.1103/PhysRevLett.108.225004}
}

@article{OhtaPOP2000,
    author = {Ohta, H. and Hamaguchi, S.},
    title = {Molecular dynamics evaluation of self-diffusion in Yukawa systems},
    journal = {Physics of Plasmas},
    volume = {7},
    number = {11},
    pages = {4506-4514},
    year = {2000},
    month = {11},
    issn = {1070-664X},
    doi = {10.1063/1.1316084},
    url = {https://doi.org/10.1063/1.1316084},
}

@article{DonkoPRE2004,
  title = {Thermal conductivity of strongly coupled Yukawa liquids},
  author = {Donk\'o, Z. and Hartmann, P.},
  journal = {Phys. Rev. E},
  volume = {69},
  issue = {1},
  pages = {016405},
  numpages = {6},
  year = {2004},
  month = {Jan},
  publisher = {American Physical Society},
  doi = {10.1103/PhysRevE.69.016405},
  url = {https://link.aps.org/doi/10.1103/PhysRevE.69.016405}
}

@book{Ferziger_Kaper,
  title={Mathematical Theory of Transport Processes in Gases.},
  author={Ferziger, J.H. and Kaper, H.G.},
  lccn={10007449},
  url={https://books.google.com/books?id=KacfcgAACAAJ},
  year={1972},
  publisher={North-Holland Publishing Company}
}

@book{Chapman_Cowling,
  title={The mathematical theory of non-uniform gases: an account of the kinetic theory of viscosity, thermal conduction and diffusion in gases},
  author={Chapman, Sydney and Cowling, Thomas George},
  year={1990},
  publisher={Cambridge university press}
}

@book{degroot_mazur,
  title={Non-equilibrium thermodynamics},
  author={De Groot, Sybren Ruurds and Mazur, Peter},
  year={2013},
  publisher={Courier Corporation}
}

@article{RynnPF1964,
    author = {Rynn, N.},
    title = {Macroscopic Transport Properties of a Fully Ionized Alkali‐Metal Plasma},
    journal = {The Physics of Fluids},
    volume = {7},
    number = {2},
    pages = {284-291},
    year = {1964},
    month = {02},
    issn = {0031-9171},
    doi = {10.1063/1.1711197},
    url = {https://doi.org/10.1063/1.1711197},
}

@article{TrintchoukPOP2003,
    author = {Trintchouk, F. and Yamada, M. and Ji, H. and Kulsrud, R. M. and Carter, T. A.},
    title = {Measurement of the transverse Spitzer resistivity during collisional magnetic reconnection},
    journal = {Physics of Plasmas},
    volume = {10},
    number = {1},
    pages = {319-322},
    year = {2003},
    month = {01},
    issn = {1070-664X},
    doi = {10.1063/1.1528612},
    url = {https://doi.org/10.1063/1.1528612},
}

@article{BretzNF1975,
doi = {10.1088/0029-5515/15/2/016},
url = {https://doi.org/10.1088/0029-5515/15/2/016},
year = {1975},
month = {apr},
publisher = {},
volume = {15},
number = {2},
pages = {313},
author = {Bretz, N. and Dimock, D.L. and Hinnov, E. and Mesrvey, E.B.},
title = {Energy balance in a low-Z high-density helium plasma in the ST Tokamak},
journal = {Nuclear Fusion}
}

@article{HawreliakJPB2004,
doi = {10.1088/0953-4075/37/7/013},
url = {https://doi.org/10.1088/0953-4075/37/7/013},
year = {2004},
month = {mar},
publisher = {},
volume = {37},
number = {7},
pages = {1541},
author = {J Hawreliak and D M Chambers and S H Glenzer and A Gouveia and R J Kingham and R S Marjoribanks and P A Pinto and O Renner and P Soundhauss and S Topping and E Wolfrum and P E Young and J S Wark},
title = {Thomson scattering measurements of heat flow in a laser-produced plasma},
journal = {Journal of Physics B: Atomic, Molecular and Optical Physics}
}

@article{HenchenPOP2019,
    author = {Henchen, R. J. and Sherlock, M. and Rozmus, W. and Katz, J. and Masson-Laborde, P. E. and Cao, D. and Palastro, J. P. and Froula, D. H.},
    title = {Measuring heat flux from collective Thomson scattering with non-Maxwellian distribution functions},
    journal = {Physics of Plasmas},
    volume = {26},
    number = {3},
    pages = {032104},
    year = {2019},
    month = {03},
    issn = {1070-664X},
    doi = {10.1063/1.5086753},
    url = {https://doi.org/10.1063/1.5086753},
}

@article{KuritsynPOP2006,
    author = {Kuritsyn, A. and Yamada, M. and Gerhardt, S. and Ji, H. and Kulsrud, R. and Ren, Y.},
    title = {Measurements of the parallel and transverse Spitzer resistivities during collisional magnetic reconnectiona)},
    journal = {Physics of Plasmas},
    volume = {13},
    number = {5},
    pages = {055703},
    year = {2006},
    month = {05},
    issn = {1070-664X},
    doi = {10.1063/1.2179416},
    url = {https://doi.org/10.1063/1.2179416},
}

@article{WhitePRL1975,
  title = {Measurement of Thermal Conductivity in a Laser-Heated Plasma},
  author = {White, M. S. and Kilkenny, J. D. and Dangor, A. E.},
  journal = {Phys. Rev. Lett.},
  volume = {35},
  issue = {8},
  pages = {524--527},
  numpages = {0},
  year = {1975},
  month = {Aug},
  publisher = {American Physical Society},
  doi = {10.1103/PhysRevLett.35.524},
  url = {https://link.aps.org/doi/10.1103/PhysRevLett.35.524}
}

@article{HohPOP1960,
    author = {Hoh, F. C. and Lehnert, B.},
    title = {Diffusion Processes in a Plasma Column in a Longitudinal Magnetic Field},
    journal = {The Physics of Fluids},
    volume = {3},
    number = {4},
    pages = {600-607},
    year = {1960},
    month = {07},
    issn = {0031-9171},
    doi = {10.1063/1.1706094},
    url = {https://doi.org/10.1063/1.1706094},
}

@article{PowersPF1965,
    author = {Powers, E. J.},
    title = {Evidence of Anomalous Diffusion in a Radio‐Frequency Discharge in a Magnetic Field},
    journal = {The Physics of Fluids},
    volume = {8},
    number = {6},
    pages = {1155-1160},
    year = {1965},
    month = {06},
    issn = {0031-9171},
    doi = {10.1063/1.1761369},
    url = {https://doi.org/10.1063/1.1761369},
}

@article{PaulikasPF1962,
    author = {Paulikas, George A. and Pyle, Robert V.},
    title = {Macroscopic Instability of the Positive Column in a Magnetic Field},
    journal = {The Physics of Fluids},
    volume = {5},
    number = {3},
    pages = {348-360},
    year = {1962},
    month = {03},
    issn = {0031-9171},
    doi = {10.1063/1.1706621},
    url = {https://doi.org/10.1063/1.1706621},
}

@inbook{bohm:1949,
	Author = {Bohm, D.},
	Editor = {Guthrie, A.},
	Publisher = {McGraw-Hill},
	Title = {The Characteristics of Electrical Discharges in Magnetic Fields},
	Year = {1949}}

@article{OttPRE2015,
  title = {Effect of correlations on heat transport in a magnetized strongly coupled plasma},
  author = {Ott, T. and Bonitz, M. and Donk\'o, Z.},
  journal = {Phys. Rev. E},
  volume = {92},
  issue = {6},
  pages = {063105},
  numpages = {8},
  year = {2015},
  month = {Dec},
  publisher = {American Physical Society},
  doi = {10.1103/PhysRevE.92.063105},
  url = {https://link.aps.org/doi/10.1103/PhysRevE.92.063105}
}

@article{SalinPOP2003,
    author = {Salin, Gwenaël and Caillol, Jean-Michel},
    title = {Equilibrium molecular dynamics simulations of the transport coefficients of the Yukawa one component plasma},
    journal = {Physics of Plasmas},
    volume = {10},
    number = {5},
    pages = {1220-1230},
    year = {2003},
    month = {05},
    issn = {1070-664X},
    doi = {10.1063/1.1566749},
    url = {https://doi.org/10.1063/1.1566749},
}

@article{StanekPOP2024,
    author = {Stanek, Lucas J. and Kononov, Alina and Hansen, Stephanie B. and Haines, Brian M. and Hu, S. X. and Knapp, Patrick F. and Murillo, Michael S. and Stanton, Liam G. and Whitley, Heather D. and Baalrud, Scott D. and Babati, Lucas J. and Baczewski, Andrew D. and Bethkenhagen, Mandy and Blanchet, Augustin and Clay, Raymond C., III and Cochrane, Kyle R. and Collins, Lee A. and Dumi, Amanda and Faussurier, Gerald and French, Martin and Johnson, Zachary A. and Karasiev, Valentin V. and Kumar, Shashikant and Lentz, Meghan K. and Melton, Cody A. and Nichols, Katarina A. and Petrov, George M. and Recoules, Vanina and Redmer, Ronald and Röpke, Gerd and Schörner, Maximilian and Shaffer, Nathaniel R. and Sharma, Vidushi and Silvestri, Luciano G. and Soubiran, François and Suryanarayana, Phanish and Tacu, Mikael and Townsend, Joshua P. and White, Alexander J.},
    title = {Review of the second charged-particle transport coefficient code comparison workshop},
    journal = {Physics of Plasmas},
    volume = {31},
    number = {5},
    pages = {052104},
    year = {2024},
    month = {05}
}

@article{FrenchPRE2022,
  title = {Electronic transport coefficients from density functional theory across the plasma plane},
  author = {French, Martin and R\"opke, Gerd and Sch\"orner, Maximilian and Bethkenhagen, Mandy and Desjarlais, Michael P. and Redmer, Ronald},
  journal = {Phys. Rev. E},
  volume = {105},
  issue = {6},
  pages = {065204},
  numpages = {9},
  year = {2022},
  month = {Jun},
  publisher = {American Physical Society},
  doi = {10.1103/PhysRevE.105.065204},
  url = {https://link.aps.org/doi/10.1103/PhysRevE.105.065204}
}

@article{WhitePRL2020,
  title = {Fast and Universal Kohn-Sham Density Functional Theory Algorithm for Warm Dense Matter to Hot Dense Plasma},
  author = {White, A. J. and Collins, L. A.},
  journal = {Phys. Rev. Lett.},
  volume = {125},
  issue = {5},
  pages = {055002},
  numpages = {6},
  year = {2020},
  month = {Jul},
  publisher = {American Physical Society},
  doi = {10.1103/PhysRevLett.125.055002},
  url = {https://link.aps.org/doi/10.1103/PhysRevLett.125.055002}
}

@article{DesjarlaisPRE2017,
  title = {Density-functional calculations of transport properties in the nondegenerate limit and the role of electron-electron scattering},
  author = {Desjarlais, Michael P. and Scullard, Christian R. and Benedict, Lorin X. and Whitley, Heather D. and Redmer, Ronald},
  journal = {Phys. Rev. E},
  volume = {95},
  issue = {3},
  pages = {033203},
  numpages = {10},
  year = {2017},
  month = {Mar},
  publisher = {American Physical Society},
  doi = {10.1103/PhysRevE.95.033203},
  url = {https://link.aps.org/doi/10.1103/PhysRevE.95.033203}
}

@article{WittePOP2018,
    author = {Witte, B. B. L. and Sperling, P. and French, M. and Recoules, V. and Glenzer, S. H. and Redmer, R.},
    title = {Observations of non-linear plasmon damping in dense plasmas},
    journal = {Physics of Plasmas},
    volume = {25},
    number = {5},
    pages = {056901},
    year = {2018},
    month = {03},
    issn = {1070-664X},
    doi = {10.1063/1.5017889}
}

@article{SjostromPRL2014,
  title = {Fast and Accurate Quantum Molecular Dynamics of Dense Plasmas Across Temperature Regimes},
  author = {Sjostrom, Travis and Daligault, J\'er\^ome},
  journal = {Phys. Rev. Lett.},
  volume = {113},
  issue = {15},
  pages = {155006},
  numpages = {5},
  year = {2014},
  month = {Oct},
  publisher = {American Physical Society},
  doi = {10.1103/PhysRevLett.113.155006},
  url = {https://link.aps.org/doi/10.1103/PhysRevLett.113.155006}
}

@article{starrett_2015,
  title = {Pseudoatom molecular dynamics},
  author = {Starrett, C. E. and Daligault, J. and Saumon, D.},
  journal = {Phys. Rev. E},
  volume = {91},
  pages = {013104},
  year = {2015},
}

@article{Baus_Hansen_1980,
title = {Statistical mechanics of simple coulomb systems},
journal = {Physics Reports},
volume = {59},
number = {1},
pages = {1-94},
year = {1980},
issn = {0370-1573},
doi = {https://doi.org/10.1016/0370-1573(80)90022-8},
url = {https://www.sciencedirect.com/science/article/pii/0370157380900228},
author = {Marc Baus and Jean-Pierre Hansen}
}

@article{LenardJMP1968,
    author = {Lenard, A. and Dyson, Freeman J.},
    title = {Stability of Matter. II},
    journal = {Journal of Mathematical Physics},
    volume = {9},
    number = {5},
    pages = {698-711},
    year = {1968},
    month = {05},
    issn = {0022-2488},
    doi = {10.1063/1.1664631},
    url = {https://doi.org/10.1063/1.1664631},
}

@article{LeVan_2025s,
  title={Plasma hydrodynamics from mean force kinetic theory},
  author={LeVan, Jarett and Baalrud, Scott D.},
  year={2025},
    journal = {Submitted Manuscript to Physics of Plasmas},
}

@article{LeePF1984,
    author = {Lee, Y. T. and More, R. M.},
    title = {An electron conductivity model for dense plasmas},
    journal = {The Physics of Fluids},
    volume = {27},
    number = {5},
    pages = {1273-1286},
    year = {1984},
    month = {05},
    issn = {0031-9171},
    doi = {10.1063/1.864744},
}

@article{KuzminPRL2002,
  title = {Numerical Simulation of Ultracold Plasmas: How Rapid Intrinsic Heating Limits the Development of Correlation},
  author = {Kuzmin, S. G. and O'Neil, T. M.},
  journal = {Phys. Rev. Lett.},
  volume = {88},
  issue = {6},
  pages = {065003},
  numpages = {4},
  year = {2002},
  month = {Jan},
  publisher = {American Physical Society},
}

@book{Reif_1965,
  address = {Tokyo},
  author = {Reif, F.},
  nota = {rk:},
  publisher = {McGraw Hill},
  title = {Fundamentals of Statistical and Thermal Physics},
  year = 1965
}

\end{document}